# Surface molecular engineering to enable processing of sulfide solid electrolytes in humid ambient air

Mengchen Liu[1,=], Jessica J. Hong[1,=], Elias Sebti[2,3,=], Ke Zhou[1,=], Shen Wang[1], Shijie Feng[4], Tyler Pennebaker[2,3], Zeyu Hui[1], Qiushi Miao[4], Ershuang Lu[5], Nimrod Harpak[1], Sicen Yu[4], Jianbin Zhou[1], Jeong Woo Oh[6], Min-Sang Song[6], Jian Luo[1,4], Raphaële J. Clément[2,3]*, Ping Liu[1,4]*

[1]Aiiso Yufeng Li Family Department of Chemical and Nano Engineering, University of California San Diego, La Jolla, CA, USA

[2]Materials Department, University of California Santa Barbara, Santa Barbara, CA, USA

[3]Materials Research Laboratory, University of California, Santa Barbara, CA, USA

[4]Program of Materials Science, University of California San Diego, La Jolla, CA, USA

[5]Department of Chemistry, University of California, San Diego, La Jolla, CA, USA

[6]LG Energy Solution, Ltd. LG Science Park, Magokjungang 10-ro, Gangseo-gu, Seoul 07796, Republic of South Korea

[=]These authors contributed equally: Mengchen Liu, Jessica J. Hong, Elias Sebti, Ke Zhou

E-mail: rclement@ucsb.edu; piliu@eng.ucsd.edu

**Abstract**

Sulfide solid-state electrolytes (SSEs) are promising candidates to realize all solid-state batteries (ASSBs) due to their superior ionic conductivity and excellent ductility. However, their hypersensitivity to moisture requires processing environments that are not compatible with today's lithium-ion battery manufacturing infrastructure. Herein, we present a reversible surface modification strategy that enables the processability of sulfide SSEs (e. g., $Li_6PS_5Cl$) under humid ambient air. We demonstrate that a long chain alkyl thiol, 1-undecanethiol, is chemically compatible with the electrolyte with negligible impact on its ion conductivity. Importantly, the thiol modification extends the amount of time that the sulfide SSE can be exposed to air with 33% relative humidity (33% RH) with limited degradation of its structure while retaining a conductivity of above 1 mS cm$^{-1}$ for up to 2 days, a more than 100-fold improvement in protection time over competing approaches. Experimental and computational results reveal that the thiol group anchors to the SSE surface, while the hydrophobic hydrocarbon tail provides protection by repelling water. The modified $Li_6PS_5Cl$ SSE maintains its function after exposure to ambient humidity when implemented in a $Li_{0.5}In||LiNi_{0.8}Co_{0.1}Mn_{0.1}O_2$ ASSB. The proposed protection strategy based on surface molecular interactions represents a major step forward towards cost-competitive and energy-efficient sulfide SSE manufacturing for ASSB applications.

**Introduction**

Owing to the safety concerns associated with the use of flammable organic electrolytes in conventional lithium-ion batteries, all solid-state batteries (ASSB) comprising a nonflammable solid electrolyte have garnered significant interest over the past decade[1–4]. Solid-state electrolytes (SSEs) also hold promise for higher interfacial stability and better mechanical properties to suppress Li dendrite penetration, opening the door to the implementation of lithium metal anodes that enables high cell energy densities[5–8]. Among SSE materials, sulfides such as $Li_{9.6}P_3S_{12}$, $Li_{10}GeP_2S_{12}$, and $Li_{9.54}Si_{1.74}P_{1.44}S_{11.7}Cl_{0.3}$ are particularly attractive due to their superionic conductivities (as high as ~$10^{-2}$ S cm$^{-1}$) and deformability[9]. Sulfide SSEs also exhibit lower Young's modulus values than oxide glasses and ceramics, which is beneficial for producing favorable interfacial contacts with electrode materials by mechanical compression[10].

An Achilles' heel of sulfide-based SSEs is their extremely poor stability in the presence of water[11,12]. Hydrolysis generates toxic H$_2$S gas together with catastrophic loss of ionic conductivity[13], requiring sulfide SSEs to be handled in an inert glove box atmosphere with exceptionally low moisture levels (-80 °C dew point) in academic studies[14]. In an industrial setting, a dry room with a dew point of < -60 °C is necessary, which is significantly more costly than the current infrastructure for lithium-ion battery manufacturing with a dew point of ~ -40 °C[15].

Recently, the degradation process of sulfide SSEs in the presence of moisture, and the main factors for the decreased ionic conductivity, have been investigated[16]. The propensity for the hydrolysis reactions of lithium argyrodites, $Li_{12-x-m}M^{m+}S_{6-x}X_x$ (*M* = P, As, Si, Ge, Sn, *X* = Cl, Br), depend on the cation $M^{m+}$ and halogen components[17]. During the hydrolysis process of $Li_6PS_5Cl$, H$_2$O tends to attack the P−S bond, thereby releasing H$_2$S by forming a P-O bond and breaking up the $PS_4^{3-}$ units that are crucial for high Li-ion conductivity[18].

There are two essential requirements for any approach to improve the moisture resistance of sulfide SSEs: 1) the protection mechanism should not significantly alter the critical properties of the material, i.e., ionic conductivity and redox stability; and 2) the protection needs to be effective for an extended period of time, on the order of hours to days, rather than minutes, commensurate with the time needed for electrolyte processing to manufacture ASSBs. In this regard, approaches based on composite formation or on modifying the composition of the SSEs themselves have been considered. Adding metal oxides (such as ZnO, $Fe_2O_3$, $Bi_2O_3$) into sulfide SSEs can suppress $H_2S$ generation through the formation of metal sulfides, but often at the expense of the ionic conductivity[19]. Alternatively, sulfide SSEs can be coated with another ionic conductor that is more resistant to moisture, to form a core-shell structure[20–23]. Sulfide SSEs can also be stabilized through substitution, either at the S or P site[24]. Partially substituting $S^{2-}$ by $O^{2-}$ to form oxysulfide compounds has led to improved air stability and wider electrochemical stability window[25]. Guided by the hard/soft acid base theory, substituting the $P^{5+}$ hard acid (small, nonpolarizable) with a $As^{5+}$, $Cu^+$ and $Sb^{5+}$ soft acid (large, polarizable) can reduce the rate of hydrolysis as the soft acids bind tightly to the $S^{2-}$ soft base, resisting the attack of the $O^{2-}$ hard base[26–28]. While the chemical stability of the electrolyte itself may improve from these substitutions, this is often at the expense of the interfacial stability against lithium metal, which jeopardizes the overall effectiveness of the ASSB. Recently, organic hydrophobic material has been shown to improve sulfide electrolyte ($Li_7P_2S_8Br_{0.5}I_{0.5}$, $Li_6PS_5Cl$) moisture stability by adsorbing on the surface through Van der Waals interaction[29,30]. Finally, regardless of the approach, reported water stability tests are usually limited to minutes of exposure time to ambient humidity conditions under 10% - 35% RH [20,29,31].

**Results**

The choice of a long-chain alkyl thiol to protect $Li_6PS_5Cl$.

In this work, we propose an alternative strategy to current surface modifiers. We note that sulfide SSEs, along with virtually all reported inorganic modifiers, are mostly hydrophilic.

While they have different abilities to resist chemical decomposition in the presence of water, they fundamentally lack the ability to repel water. On the other hand, previously reported hydrophobic organic modifier has weak Van der Waals interaction with the electrolyte. Here, we introduce an amphiphilic molecule, 1-undecanethiol (UDSH), that can chemically adsorb onto the $Li_6PS_5Cl$ (LPSC) SSE surface (labeled as UDSH@LPSC) to form a hydrophobic shield (Fig. 1). Long-chain alkane thiol is a classic example for the formation of self-assembled monolayers (SAM) on a variety of substrates[32]. They are ideally suited for the protection of sulfide solid electrolytes. Sulfide SSEs are highly susceptible to nucleophilic attacks from polar organic solvents (e.g., diethylene glycol dimethyl ether or propylene carbonate)[33,34]. Very few functional groups that chemically interact with the surface of LPSC without altering the crystal structure have been identified. Toluene, an unsaturated hydrocarbon, is thus the most common choice. Their interaction with the LPSC is limited to Van der Waals forces, and not polar enough to enable chemisorption[35]. These findings highlight the need to tailor the chemistry of the head group of the surface modifier to enable both stabilization of and attachment to the sulfide SSE surface. A thiol head group is an ideal choice. It cannot launch nucleophilic attack on the P-S bond while offering the possibility of forming S-S bonds with the surface of LPSC without altering its crystal structure and its ion conduction property[29]. We note that a thiol ((trimethylsilyl)ethanethiol) has been employed to enhance cathode/electrolyte interface stability[36]. The hydrocarbon tail of UDSH is also essential: it provides the needed hydrophobicity to mitigate access of water to the electrolyte surface. It is rather fortuitus that the most common SAM molecule happens to be the most suitable for sulfide protection although the rationales are unrelated: thiol-based SAM was overwhelmingly applied to metal surface where the formation of the S-metal strong chemisorption scheme is the main motivation. In contrast, their ability to form S-S bond with LPSC is crucial for their application in our case.

In order to gain insights into the UDSH water repelling ability upon prolonged air exposure, we examined the interaction of the UDSH with humid air in the absence of LPSC.

Supplementary Fig. 1 shows the 300-1900 cm$^{-1}$ region of Raman spectra recorded on an UDSH solution and upon exposure to air with 33% RH for up to 3 days, to monitor changes in the C−S stretching frequency. UDSH maintains the C−S bond peak at 639 cm$^{-1}$, and there is no evidence for S-S bond formation at 450-495 cm$^{-1}$, which suggests that UDSH molecules retain their structural integrity in ambient air[37]. $^1$H Nuclear magnetic resonance (NMR) spectra, shown in Supplementary Fig. 2, were also collected on the same series of UDSH samples and exhibit six $^1$H resonances that are assigned to the six $^1$H local environments present in the UDSH molecule[38]. With prolonged exposure, $^1$H signals exhibit constant isotropic shifts, no appearance of H$_2$O peaks, and minimal changes in signal intensities. This implies that the UDSH molecule's structure remains unchanged, and H$_2$O diffusion within UDSH is exceptionally slow.

The introduction of a small molecule into the system leads to good retention of the ionic conductivity of the SSE, unlike previously reported polymeric surface modifier[39]. The protection mechanism is highly effective: the UDSH@LPSC maintains the ionic conductivity to above 1 mS cm$^{-1}$ for up to 2 days of exposure (33% RH). Even after 3 days of exposure, UDSH@LPSC powder maintains its original color and crystallinity, while the control material suffered catastrophic loss of conductivity along with clear discoloration (Fig. 1).

Chemical compatibility between UDSH and LPSC

To probe the chemical interaction between the LPSC and the UDSH, we first conducted density functional theory (DFT) calculations of the adsorption energies of undecane (UDCH) (Supplementary Fig. 3b), which does not contain the thiol head group, and UDSH (Supplementary Fig. 3c), which contains the thiol functional group, onto the surface of LPSC (Supplementary Fig. 3a). Fig. 2a demonstrates that both organic compounds can adsorb onto the LPSC surface, but UDSH exhibits a much higher propensity to adsorb onto the surface (- 3.821 eV) as compared to UDCH (-1.172 eV). Furthermore, UDSH can indeed form S-S bonds with LPSC, in line with our expectations regarding the significance of the thiol head group.

We then prepared UDSH modified LPSC (UDSH@LPSC) by mixing desired amounts of the materials using a planetary centrifugal mixer (Supplementary Fig. 4). This mixture was then subjected to vacuum drying at 80°C for 2 hours to remove residual unbonded UDSH on LPSC surface. The compatibility of UDSH with LPSC was evaluated by monitoring potential changes in crystal structure and ionic conductivity of UDSH@LPSC. Rietveld refinement of an X-ray Diffraction (XRD) pattern obtained on the UDSH@LPSC sample (Supplementary Fig. 5) indicates that the argyrodite structure remains intact. $^1$H spin echo NMR spectra, shown in Fig. 2b, were collected on a pure UDSH solution and on the UDSH@LPSC sample to investigate the nature of the interactions between UDSH and the LPSC surface. Six $^1$H resonances can be resolved in the spectrum obtained on the UDSH solution, that are assigned to the six $^1$H local environments, labeled (a-f), in UDSH molecule. This assignment is based on predicted $^1$H shifts from the ACD/HNMR Predictor software package[38], and from the observed integrated signal intensities and expected $^1$H site multiplicities. In UDSH@LPSC, all the UDSH resonances remain visible, but the (a) and (b) resonances are shifted downfield (to more positive ppm frequencies). The (a) and (b) resonances correspond to $^1$H nuclei bonded to carbons one and two bonds away from the thiol end group, respectively. In NMR, the isotropic shift of a nucleus depends on the degree of shielding from the static magnetic field provided by the local electron cloud and is therefore a sensitive probe of local structure changes. Here, changes in the isotropic shifts of the (a) and (b) resonances suggest that the UDSH molecules bond to the LPSC surface at the thiol end. The downfield shift of the (a) resonance (0.2 ppm) is greater than that of the (b) resonance (0.08 ppm), in line with the greater change in the electron cloud around the $^1$H nucleus closest to the thiol end. While the seeming disappearance of the (d) resonance assigned to the thiol group $^1$H in the UDSH@LPSC spectrum could provide further evidence for thiol end attachment to the LPSC surface, this signal overlaps with the (c) and (e) resonances, and the presence of residual, free UDSH molecules in the UDSH@LPSC sample makes it difficult to correlate changes in (d) signal intensity with attachment chemistry. An upfield shift is observed for the (f) resonance (0.08 ppm) that corresponds to the UDSH methyl chain end. We tentatively attribute this slight upfield shift to weak Van der Waals interactions between the methyl end of the UDSH molecules and the LPSC surface. The broad $^1$H resonance observed between 0.5-1 ppm in the UDSH@LPSC spectrum in Fig. 2b corresponds to protonated impurities in the LPSC SSE, as confirmed by the $^1$H spectrum obtained on pure LPSC and shown in Supplementary Fig. 6. We also find that subjecting the UDSH@LPSC sample to a vacuum drying procedure at 80°C for 2 hours do not completely remove UDSH, as indicated by the UDSH signals remaining in the corresponding $^1$H NMR spectrum (Supplementary Fig. 6), with signal integration suggesting about 4-5 wt% of UDSH in the sample. However, almost all the UDSH modifiers can be volatilized after an additional heat treatment at 300°C if desired (Supplementary Fig. 6). Further details on the quantification procedure are provided in Supplementary Fig. 6, Supplementary Note 1, and representative fits of the corresponding spectra are provided in Supplementary Fig. 7.

The bonding interaction between UDSH and LPSC is further investigated using surface-sensitive X-ray photoelectron spectroscopy (XPS) analysis, with results presented in Fig. 2c-d. LPSC and UDSH@LPSC display a P-S bond from $PS_4^{3-}$ at 161.5 eV in the S *2p* spectrum, and at 131.9 eV in the P *2p* spectrum[40]. For LPSC, the signals for Li-S at 160 eV (Fig. 2c) and P-O at 133.5 eV (Fig. 2d) are attributed to surface decomposition products such as $Li_2S$ and $PO_xS_y$[40]. Within UDSH@LPSC, the increased intensity of the S-H peak at 160 eV, which overlaps with Li-S, alongside the concurrent decrease in the P-O peak at 133.5 eV, suggest the creation of a thin UDSH film on the LPSC surface. The increased intensity of the S-S peak at 164.5 eV on the UDSH@LPSC sample (Fig. 2c), can be attributed to bridging sulfur atoms[41]. This provides additional evidence for the establishment of S-S bonds between UDSH molecules and S atoms at the surface of LPSC. The integrity of the P-S bond at 131.9eV in Fig. 2d indicates the good compatibility of UDSH and LPSC. These findings corroborate the existence of S-S interaction between UDSH and LPSC, aligning well with the DFT calculation and NMR results.

Cryogenic transmission electron microscopy (cryo-TEM) images of pristine LPSC and UDSH@LPSC samples reveal the nanoscale distribution of crystalline and amorphous regions, as shown in Fig. 2e-f. The presence of an UDSH layer on the surface of LPSC does not alter its bulk crystal structure, as indicated by the similar lattice spacing measured on UDSH@LPSC (of 0.182 nm, Fig. 2f), as compared to that measured on pristine LPSC (0.185 nm, Fig. 2e). This observation validates the compatibility of UDSH with LPSC. An amorphous layer of ~ 3 nm in thickness is present at the surface of the LPSC particle in the UDSH@LPSC sample, which we attribute to the UDSH adsorption. The presence of UDSH on the LPSC is also corroborated by Energy Dispersive X-ray Spectrometry (EDXS) results which confirm a uniform carbon-rich layer on the surface of LPSC (Supplementary Fig. 8-9, Supplementary Table 1). The structure of self-assembled alkyl thiol layers has been traditionally characterized by atomic force microscopy (AFM) conducted on flat substrates[32]. That is not feasible on a powder sample with rough surfaces. To circumvent this difficulty, we prepared a thin film LPSC layer by spin coating from an anhydrous ethanol solution followed by thermal annealing (Supplementary Fig. 10a-b). XRD spectra in Supplementary Fig. 10c confirm the crystal structure to be consistent with the argyrodite structure. Two thiols, UDSH and HDSH (hexadecane thiol, $C_{16}H_{33}SH$), were adsorbed onto the LPSC film surface. AFM images show the formation of SAM-type layers for UDSH, while the HDSH layer appears to be far more disorganized with the hydrocarbon chain lying parallel to the substrates (Supplementary Fig. 11-12). These observations indicate that UDSH is a preferred protection agent than those with longer chains by forming a higher quality surface coating. As will be shown later, the chain length has a significant effect on their efficacy as protection agents.

The combined bulk and grain boundary resistance of UDSH@LPSC exhibits a slight increase (Fig. 2g) as compared to pristine LPSC. This modest increase can be attributed to the presence of 4-5 wt% of UDSH adsorbent that remains after the 80°C drying procedure. Li-ion conductivities are 2.5±0.05 mS cm$^{-1}$ and 2.1±0.2 mS cm$^{-1}$, respectively, based on the average of 3 independent measurements (errors correspond to one standard deviation). The modest effect of the UDSH surface modifier on the conductivity of LPSC is highly unusual. For example, when subjecting sulfide SSEs to other polar solvents, such as diethylene glycol dimethyl ether or propylene carbonate, a two order of magnitude decrease in ionic conductivity is typically observed due to nucleophilic attack from electron donor from solvent molecules to the SSE species[33]. In the case of propylene carbonate, the resulting mixture becomes sticky and exhibits ionic conductivity too low to be measured[33].

To examine the role of the UDSH molecule in influencing interfacial ion transport, we conducted 2D NMR using the exchange spectroscopy (EXSY) method which measures the Li ion exchange rate between two different solid electrolytes (Supplementary Fig. 13). Here, LPSC with or without the UDSH coating was mixed with Li$_2$ZrCl$_6$ (LZC). The diagonal peaks represent lithium nuclei which remain in the same environment during the given mixing time, while cross-peaks arise from the signal of nuclei which have hopped between environments. The cross-peak intensity ratio is 0.16 for pristine LPSC with LZC and 0.22 for UDSH@LPSC with LZC. The higher ratio in the UDSH@LPSC means that more lithium is hopping between the LPSC and LZC during the provided mixing time, thus proving that the UDSH coating does not prevent diffusion between LPSC and other nearby environments. More detailed analysis is provided in Supplementary Fig. 13, Supplementary Note 2.

Effectiveness of protection against moisture.

The ultimate goal of this work is to drastically enhance the chemical stability of LPSC by protecting it against hydrolysis. LPSC and UDSH@LPSC were evaluated after exposure to humid ambient air from 5 hours and up to 3 days. The samples are labeled as "LPSC time air", e.g., LPSC 5H air refers to a sample that has been exposed for 5 hours.

To evaluate the air stability of various LPSC samples in a quantitative manner and under specific humidity conditions, a humidifier was used ensuring a consistent initial humidity of 33% RH. 600mg LPSC or a mixture of 600mg LPSC and 120mg UDSH, powder instead of pellet, was placed inside the container. A thermo-hydrometer was positioned near the sample to monitor changes in atmospheric humidity levels over time. Supplementary Fig. 14a shows the evolution of the relative humidity (RH) in the container as a function of exposure time (t). After 72 hours (3 days) of exposure, the relative humidity inside the LPSC container decreases from 33% to 7%. In contrast, the relative humidity within the UDSH@LPSC container decreases from 33% to 13%. The first derivative of the relative humidity with respect to time provides an estimate of the rate of water uptake by the LPSC. The results, shown in Supplementary Fig. 14b, reveal a significantly reduced rate of water uptake for UDSH@LPSC, further confirming the effectiveness of the UDSH surface modifier at mitigating moisture reactivity. Concurrently, the production of $H_2S$ was also reduced by 5 times after 3 days (Supplementary Fig. 15). LPSC showed continuous $H_2S$ generation during exposure, while UDSH@LPSC exhibited a much slower $H_2S$ generation rate. The generation of $H_2S$ should be closely related to structural degradation and loss of conductivity, which indicates that alkylthiol protections slows down both. The above air exposed samples were subjected to a vacuum drying procedure at 80°C for 2 hours prior to following characterizations. We note that the 2-hour heat treatment at 80°C has no discernible effect on conductivity (Supplementary Notes 3). The complete removal of UDSH by a 3-hour heat treatment at 300°C also indicates no degradation. (Supplementary Fig. 16, 17)

    The ionic conductivity of electrolyte pellets prepared by cold pressing is plotted in Fig. 3a, based on the average of 3 independent measurements (errors correspond to one standard deviation). When UDSH absorbent is applied, the UDSH@LPSC ionic conductivity slightly decreases from 2.5±0.05 mS cm$^{-1}$ to 2.1±0.2 mS cm$^{-1}$, as shown in Fig. 3a. UDSH@LPSC

exhibits excellent stability after air exposure, based on the very small change of its ionic conductivity from 1.8±0.10 mS cm$^{-1}$ to 0.8±0.27 mS cm$^{-1}$ after 5 hours to 3 days (Fig. 3a). In contrast, LPSC without the UDSH coating experiences a significant drop in ionic conductivity to 0.19±0.003 mS cm$^{-1}$ in 5 hours and to 8*10$^{-4}$ ±4*10$^{-5}$ mS cm$^{-1}$ after 3 days of air exposure (Fig. 3a). The rapid decrease in ionic conductivity of LPSC within the first 5 hours is consistent with the findings in previous reports[20,40].

The change in ionic conductivity of LPSC during exposure is a function of both humidity and time. When the amount of LPSC powder is reduced from 600 mg (Fig. 3a) to 200 mg (Supplementary Fig. 18), thereby increasing the H$_2$O to LPSC ratio, the relative effectiveness of the UDSH protection becomes significantly more pronounced. UDSH@LPSC exhibits an ionic conductivity of 0.4±0.01 mS cm$^{-1}$ that is two orders of magnitude higher than LPSC (0.003±1.5*10$^{-4}$ mS cm$^{-1}$) after 1 day of air exposure. Further, we examined the moisture stability of LPSC at UDSH:LPSC ratios of 1:5, 1:7, 1:9 and 1:24 with the results shown in the Supplementary Fig. 19. The sample with a ratio of 1:24 is prepared by vacuum drying when all remaining UDSH (4 wt%) is chemically adsorbed. In ambient air with an RH of 33%, the 1:24 sample maintained an ionic conductivity of 1.0 mS cm$^{-1}$ after 5 hours, five times higher than that observed for LPSC without UDSH protection. There is indeed a correlation between UDSH:LPSC ratio and protection efficacy, with much longer protection for increased UDSH amount. This indicates that both the surface SAM layer and the hydrophobic UDSH liquid layer mitigate the diffusion of water to LPSC surface. The efficacy of the surface layer depends on its structure. On a rough particle surface such as LPSC, it is difficult to form compact, defect free SAM layers[32]. NMR results indicate that the hydrocarbon tail also has interactions with the particle surface which is Van der Wall in nature.

To put our results in context, Fig. 3b-c summarizes previously reported stability data collected on modified LPSC electrolytes after air/moisture exposure[20,31,42–46]. UDSH protection extends the LPSC protection time from minutes to days (Fig. 3b). The conductivity loss rate is in general ~ 3 orders of magnitude lower (Fig. 3c). Our approach could enable sulfide SSE processing under standard atmospheric conditions with no additional humidity controls, dramatically driving down manufacturing costs. Needless to say, their stability in a standard dry room with a dew point of -40 °C is fully expected. Moreover, we have found that UDSH can dissolve the common binder for SSE, hydrogenated nitrile butadiene rubber (HNBR), with a solubility of 10 wt% at room temperature (20±2°C) (Supplementary Fig. 20a-b). UDSH can thus serve as both a processing solvent and a protection agent, enabling processing of solid electrolyte films in less controlled dry room environment (Supplementary Fig. 20c-f, Supplementary Note 4).

Structural evolution during air exposure.

The progression of materials structural changes following exposure to air (33%RH) was systematically examined. The XRD patterns in Fig. 4a indicate that, following 1 day of exposure in 33% RH air, no new phases are formed in the UDSH@LPSC sample, while new peaks at 25° and 35° with very low intensities start to appear after 3 days of exposure. The basic structure of LPSC remains unaltered, suggesting the effective role of UDSH as a protective agent. On the contrary, the XRD pattern of LPSC after being exposed for 3 days in 33% RH air exhibits major $Li_2S$ and LiCl decomposition phases, indicated by the peaks at 25° and 35° [40]. The complete XRD dataset is provided in Supplementary Fig. 21.

Raman spectroscopy was used to monitor the evolution of the LPSC structure. Here, we monitor changes in the peak at 424.5 cm$^{-1}$ (Fig. 4b), corresponding to $PS_4^{3-}$ units, a key structural component that is responsible for the excellent ionic conduction of sulfide SSEs[47]. The peak shifts to lower wavenumbers as the material degrades, which is attributed to the decomposition of P-S bonds into P-O bonds, potentially forming oxy-thiophosphate species, although the exact speciation requires further investigation[47]. The spectrum obtained on the UDSH@LPSC 1D air sample exhibits no change in the 424.5 cm$^{-1}$ peak, while the

UDSH@LPSC 3D air sample shows a very small shift of the same peak when 16.2% of $PS_4^{3-}$ unit is degraded. In contrast, the LPSC 1D air sample exhibits a substantial peak shift and signifies 41.8% degradation of $PS_4^{3-}$ unit. These results agree with the XRD findings indicating that the bulk structure only starts to degrade at about 3 days of humid air exposure for the UDSH@LPSC SSE.

The progression of materials changes was further monitored using XPS and solid-state NMR (ssNMR), with results shown in Fig. 4c-e. Upon examining the LPSC 1D air and LPSC 3D air samples, S-O peaks from $SO_x^{2-}$ at 167 eV appear (Fig. 4c) and the intensity of the P-O peak from oxysulfides ($PO_xS_y$) increases at 132.5 eV (Fig. 4d), indicating that hydrolysis has occurred, causing $S^{2-}$ oxidation. Conversely, S-O peak is absent from UDSH@LPSC 1D air and UDSH@LPSC 3D air samples (Fig. 4c), and the intensity of P-O peaks remains unchanged (Fig. 4d) when compared to the UDSH@LPSC (Fig. 2d). The S-S bond intensities in UDSH@LPSC 1D and UDSH@LPSC 3D are almost identical to UDSH@LPSC. We credit the bonding between UDSH and LPSC for protecting the surface from oxidation, while the hydrolysis reaction is significantly reduced so that the critical $PS_4^{3-}$ units are effectively preserved.

LPSC and UDSH@LPSC samples were also investigated with $^{31}P$ and $^6Li$ ssNMR before and after humid ambient air exposure, with results shown in Fig. 4e and Supplementary Fig. 22, respectively. Because ssNMR is element-specific, and sensitive to both crystalline and amorphous phases, it can selectively probe $^{31}P$ and $^6Li$ nuclei within the LPSC electrolyte. No clear differences are observed between the $^{31}P$ spectra obtained on the LPSC and UDSH@LPSC samples, signifying that the LPSC bulk structure is unchanged upon mixing with UDSH. The $^{31}P$ ssNMR spectrum collected on the UDSH@LPSC 1D air sample is nearly identical to that obtained on LPSC, indicating that the coating protects the LPSC structure from degradation. Even after 3 days in air, most of the $^{31}P$ spectral features attributed to pristine LPSC are retained in the UDSH@LPSC sample. Nevertheless, the onset of degradation is suggested by the appearance of new resonances at 82.9 and 88.7 ppm, as well as a third resonance at 86.8 ppm assigned to isolated $PS_4^{3-}$ tetrahedra in $\beta$-$Li_3PS_4$-like environments[48]. Without any protection, air exposure leads to significant degradation of the LPSC bulk structure, reflected by the loss of the broad pristine LPSC resonance and the appearance of sharp resonances at 82.9 and 88.7 ppm, assigned to $POS_3^{3-}$ tetrahedral units and isolated $PS_4^{3-}$ tetrahedra in $Li_3PS_4$-like environments, respectively[40,48]. Interestingly, air exposure of the UDSH@LPSC samples appears to lead to the formation of more $Li_3PS_4$-like isolated $PS_4^{3-}$ tetrahedra, as evidenced by the greater intensity of the resonances at 86.8 ppm and 88.7 ppm in the spectrum collected on the UDSH@LPSC 3D air sample relative to that obtained on the LPSC 3D air sample. Overall, $^{31}P$ and $^6Li$ ssNMR results clearly demonstrate that the UDSH modifying delays the onset of degradation and largely preserves the LPSC structure after up to 3 days of exposure to ambient air. Finally, $^1H$ spectra collected on UDSH@LPSC and UDSH@LPSC 1D air samples (Supplementary Fig. 23) suggest that the amount of UDSH adsorbed onto the LPSC surface decreases with air exposure time (presumably due to evaporation), as indicated by a decrease in the broad peak near (a) (~3.5 ppm) decreases, which could partly explain the reduced efficiency of the UDSH coating with increasing exposure time.

Discussion on the protection mechanisms.

While we have established the effectiveness of UDSH in protecting LPSC through the formation of a surface layer, its exact working mechanism deserves further examination. In this regard, we first investigated a variety of hydrophobic materials with different functional groups to see whether hydrophobicity itself is sufficient for protection. The list includes petroleum wax ($C_nH_{2n+2}$, n=20-40), perfluoropolyether oil (PFPE, $C_{50}F_{102}O_{16}$) (Supplementary Fig. 24), undecane (UDCH) and hexadecane (HECH) (Fig. 5a). The weight ratio of the hydrophobic material to LPSC is 1:5. LPSC protected by petroleum wax or PFPE oil suffered significant changes in crystal structure after just one day of exposure in 33% RH humid air (Supplementary

Fig. 24a-b). As a control for UDSH, we also measured the air protection effect of UDCH and HECH which do not have the thiol head group. The ionic conductivity of UDCH or HECH coated LPSC dropped significantly from 1.8 to 0.7 mS cm$^{-1}$ and 1.6 to 0.2 mS cm$^{-1}$ after 1 day of 33%RH air exposure, respectively (Fig. 5a). $^6$Li and $^{31}$P spectra were also collected on a series of samples modified with UDCH instead of UDSH both before and after exposure to 33%RH air for one or three days. Coating with UDCH (with a methyl rather than a thiol end group) does not protect the LPSC SSE against hydrolysis, and significant changes are observed in the $^6$Li and $^{31}$P resonances of the UDCH@LPSC 1D air sample (Supplementary Fig. 25-26). The results shown in Supplementary Fig. 24-26 again indicate that the effectiveness of the UDSH is not just the result of having the material being covered by excessive amount of hydrophobic material. The presence of the thiol headgroup is essential. Finally, we note that polar head group needs to be "oxygen" free to preserve conductivity. The mixing of undecanol with LPSC (labeled as UDOH@LPSC) leads to > 4 magnitude of increase in LPSC resistance even before air exposure as shown in Supplementary Fig. 24c.

We also examined the role of hydrocarbon tail length since it is known to affect the quality of the self assembled monolayer[32]. Octanethiol (labeld as OCSH, which has 8 -CH$_2$-), undecanethiol (labeled as UDSH, which has 11 -CH$_2$-), dodecanethiol (labeled as DOSH, which has 12 -CH$_2$-), tetradecanethiol (labeled as TESH, which has 14 -CH$_2$-) and hexadecanethiol (labeled as HESH, which has 16 -CH$_2$-) are evaluted as protection agent for LPSC (Fig. 5b). While all thiols show some benefits when the liquid thiol thickness is consistent, UDSH offers the most effective protection in terms of ionic conductivity, based on the average of 2 independent measurements (errors correspond to one standard deviation), likely due to the better quality of the surface protection layer. As an example, the more effective protection of UDSH@LPSC than HESH@LPSC after air exposure in Fig. 5b is consistent with the higher surface layer quality and more vertical chain orientation of UDSH from AFM results (Supplementary Fig. 11-12).[32]

The confirmed air stability of UDSH, and its ability to form S-S bonds with the LPSC surface, make it a powerful, hydrophobic protecting agent. Specifically, UDSH forms a layer that impedes water from reaching the LPSC surface. This protective layer prevents LPSC from reacting with ambient moisture for up to several days. The slow degradation of UDSH@LPSC upon extended air exposure likely involves water penetration through defects that bypass the UDSH protective layer, or through UDSH evaporation that leaves unbonded LPSC surface exposed.[49]

Enhanced Electrochemical Performance.

We next examine whether LPSC and UDSH@LPSC SSEs exposed to air for 1 day can maintain their electrochemical stabilities in Li symmetric and full cells. Li||Li symmetric cells are commonly used to evaluate the interfacial stability of SSEs with lithium metal. Critical current density (CCD) is a metric to measure how well the SSE resists shorting due to lithium dendrite growth[6]. Supplementary Fig. 27 shows that the CCD value of UDSH@LPSC 1D air remains unchanged compared to pristine LPSC, with a value of 0.6 mA cm$^{-2}$, while the CCD value of LPSC 1D air decreases from 0.6 to 0.3 mA cm$^{-2}$ at room temperature (20±2°C). Thus, in UDSH@LPSC 1D air, the surface modification along with the air exposure do not appear to change the interactions between LPSC and Li. Furthermore, Fig. 6 illustrates the galvanostatic cycling performance of Li$_{0.5}$In||LiNi$_{0.8}$Co$_{0.1}$Mn$_{0.1}$O$_2$ with a capacity loading of 1.5 mAh cm$^{-2}$. As compared to a base line cell using a pristine LPSC electrolyte (Supplementary Fig. 28), the cell fabricated with LPSC after 1 day of air exposure (Fig. 6a-b) exhibits ~80 mV decreases in cell voltage as well as a lower discharge capacity of 128±15 mAh g$^{-1}$ due to the increased electrolyte resistance after the formation cycle. In contrast, the cell fabricated with

UDSH@LPSC after 1 day of air exposure (Fig. 6c-d) demonstrates similar capacity of 152±6 mAh g$^{-1}$ and 82±1.6% capacity retention even after 50 cycles as the pristine LPSC full cell (Supplementary Fig. 28), based on the average of 2 independent measurements (errors correspond to one standard deviation). This observation serves as strong evidence of the remarkable protective effect of UDSH on LPSC, particularly in mitigating the adverse effects of moisture exposure.

We have shown that a long chain thiol, 1-undecanethiol, can be used to effectively protect the Li$_6$PS$_5$Cl solid electrolyte from humid air exposure. 1-undecanethiol is found to be chemically compatible with Li$_6$PS$_5$Cl, with negligible impact on its conductivity, and can be mostly removed with a vacuum drying process. Structural analysis shows that the -SH end group forms S-S bonds with S atoms at the surface of the solid electrolyte without disrupting the P-S network, which is essential for ion conductivity. The hydrocarbon tail of the thiol forms a hydrophobic layer that effectively repels water. The UDSH surface modification is found to be both effective at maintaining the bulk structure and ionic conductivity of Li$_6$PS$_5$Cl, even upon extended exposure to humid air. The UDSH-modified solid electrolyte maintains an ionic conductivity > 1 mS cm$^{-1}$ for 2 days when exposed to air with a relative humidity of 33%. These exceptional outcomes constitute a substantial advancement in terms of moisture protection time, outperforming prior work in this area by two orders of magnitude. With the 1-undecanethiol protection, the electrolyte shows comparable performance in solid state batteries even after 1 day of air exposure. While practical implementation of using thiol requires scale up and cost reduction, our results demonstrate their potential for preserving the extremely air-sensitive sulfide solid electrolytes outside of a glovebox or in a dry room. They can also serve as the solvents for polymer binders during electrolyte processing. The introduction of the long-chain thiol compound paves the way for more practical, efficient, and scalable sulfide solid state battery production processes that are compatible with today's lithium-ion battery manufacturing infrastructure.

**Methods**

DFT calculation: The density functional theory (DFT) calculations were conducted using the classic VASP code[50]. The Perdew-Burke-Ernzerhof generalized gradient approximation (PBE-GGA) [51] was used for full relaxation and calculations. A kinetic energy cut-off of 500 eV was used when applying the projector augmented-wave pseudopotential (PAW)[52]. The vacuum layer was fixed to 15 Å to prevent interaction between LPSC and two organic materials (UDCH and UDSH). The Brillouin-zone (BZ) of the unit cell was sampled using a Γ-centered 3 × 3 × 1 Monkhorst-Pack k-point mesh. All atomic positions were fully relaxed until the total energy difference was $1 \times 10^{-5}$ eV and atomic force difference was 0.03 eV/Å.

$Li_6PS_5Cl$ surface engineering: LPSC (>95%) was purchased from NEI corporation and was used as received, UDSH (98%), UDCH (≥99%), OCSH (>98.5%), DOSH (>98%), TESH (≥98%) and HESH (99%) was purchased from Sigma Aldrich and was used as received. The UDSH@LPSC samples were prepared by planetary mixing LPSC and UDSH with a 5:1 weight ratio in a 12 ml plastic vial using 6 of zirconium balls (2mm diameter) with a planetary mixer (Thinky) for 20 minutes at 1800 rpm. All samples were handled in a glovebox with oxygen and water levels below 0.5 ppm.

$Li_6PS_5Cl$ Air exposure: In our home designed air exposure set up, we used compressed air from Airgas AI UZ300 with 76.5 - 80.5% nitrogen and 19.5-23.5% oxygen. The air flow for humidification is set at 100 linear feet per minute (fpm) until the humidity inside the container stabilizes at 33% relative humidity (RH). The water container, with a volume of 400 ml, contains a 200 ml volume of water and 200mL of headspace. The 12-ml plastic vial (Thinky container) with the mixture of 600 mg of LPSC powder and 120 mg of UDSH, or 600mg of LPSC powder instead of pellets are put into a 16 oz jar for air exposure with initial humidity 33%RH at room temperature (20±2°C). The exposure times are 5 hours, 1 day, 2 days, and 3 days. Real-time monitoring of humidity during LPSC and UDSH@LPSC exposure are accomplished using a Govee Thermometer Hygrometer positioned in close proximity to the powder.

After air exposure, the LPSC and UDSH@LPSC samples were taken into the vacuum oven to dry excess solvent and moisture of the surface of the materials at 80 °C for 2 hours with a heating ramp of 1 °C/min. The dried samples were then moved into the glovebox for further tests. Note that the complete removal of all UDSH residues can be achieved through a 300°C heating process where the crystal structure and conductivity are consistent with the pristine LPSC. However, we have found that the 80ºC drying step is sufficient to remove most of the UDSH without detrimental effect on conductivity and stability with Li.

LPSC film processing using UDSH as a solvent: A solution comprising 5 wt% HNBR (Sigma Aldrich) in UDSH was prepared at room temperature (20±2°C). Subsequently, 600 mg of LPSC and 500 mg HNBR/UDSH solution was mixed in a 12 ml plastic vial using zirconium balls with a planetary mixer (Thinky) for 20 minutes at 1800 rpm, yielding a uniform wet paste. This paste was then cast into a uniform film using doctor blade within an Argon filled glove box or a dry room with a dew point of -28ºC. The thus fabricated film was dried in vacuum at 80ºC for 6 hours.

LPSC Thin Film Preparation and AFM Characterization: LPSC thin films were deposited by spin coating onto glass substrates. LPSC and anhydrous ethanol (≥99.5%, Sigma Aldrich) were used as a starting material and solvent, respectively. The concentration of LPSC was 0.1 wt%.

The glass substrate was pre-cleaned sequentially by detergent water, DI water, anhydrous ethanol and acetone (≥99.5%, Sigma Aldrich) for 10 min each by using a Digital Pro ultrasonic cleaner and then dried. 200 μl of the coating solution was dropped onto a glass substrate, which was rotated at 1000 rpm for 60s by using an MTI VTC-50A spin coater. The deposited film was heated at 180ºC for 6 hours in an Ar filled glovebox to evaporate the solvent. To form self-assembled layers of thiol on LPSC, 20 mg of UDSH or HESH was deposited onto the surface of the thin film LPSC within an argon-filled glovebox, followed by vacuum drying at 80°C for 2 hours to remove unbonded thiol. The surface topography of the LPSC thin film was characterized using a Park NX20 Atomic Force Microscope (AFM). All topographical images were captured under ambient air conditions and at room temperature (20±2°C), ensuring a brief exposure time of less than 5 minutes, utilizing the tapping mode for measurement. Silicon cantilevers, featuring a resonance frequency of 300 kHz and a force constant of 26 N/m, were employed for this purpose. Image processing and analysis were conducted using the Park Systems XEI software, facilitating the detailed examination of the thin film's surface characteristics.

Materials Characterization: Raman spectroscopy (Renishaw™ inVia™) measurements were taken from the range 300 ~ 2500 $cm^{-1}$ at an excitation wavelength of 785 nm. All powders were hand ground and prepared in a glovebox covered with Kapton tape to prevent air exposure during the measurements. Liquid UDSH samples before and after air exposure were analyzed via $^1$H NMR spectroscopy. To prepare the sample, 100 μl of UDSH or air exposed UDSH specimen was mixed with 1000 μl of chloroform D (99%, Sigma Aldrich), and then transferred to an NMR sample tube. $^1$H NMR is performed with a 500 MHz spectrometer (ECA500, JEOL). Powder XRD (D8 DISCOVERY, Bruker) was measured from 10° – 80 ° at a scan rate of 2° / min. The sample was sealed using a Kapton tape. In the cryo-TEM imaging, the specimens were placed onto a copper grid and affixed to a dual-tilt cryo-TEM transfer holder within a liquid nitrogen environment. Cryo-TEM images were captured utilizing a JEM-2100F electron microscope operating at an acceleration voltage of 200 kV. XPS analyses were conducted using an AXIS Supra XPS instrument by Kratos Analytical. The XPS data was acquired utilizing a monochromatized Al Kα source emitting at 1,486.7 eV with samples under an ultra-high vacuum environment of $10^{-8}$ Torr. The samples were transferred from a nitrogen-filled glovebox to avoid further air exposure. Prior to analysis, a 10 keV Ar (1000 atom) cluster source was applied for a duration of 60 seconds for surface cleaning. All XPS spectra were calibrated with the adventitious C 1s peak at 284.6 eV and subsequently analyzed using the CasaXPS software.

Solid State NMR: All solid-state NMR spectra were acquired at 18.8 T (800 MHz for 1H) on a Bruker Ultrashield Plus standard bore magnet equipped with an Avance III console. High resolution $^1$H and $^{31}$P solid-state NMR spectra were obtained using a 1.3 mm HX MAS probe and at a magic angle spinning (MAS) speed of 60 kHz. The powder samples were packed into zirconia rotors sealed with Vespel caps under argon. A flow of dry $N_2$ gas (2000 l/hr) was used to control the temperature of rotor and protect the sample from moisture exposure. A rotor synchronized spin-echo pulse sequence (90°-$\tau_r$-180°-$\tau_r$-acquisition) were used to obtain data. For $^1$H, 90° and 180° flip angles of 1.6 μs and 3.2 μs, respectively, at 75 W were used. For $^{31}$P, 90° and 180° flip angles of 1.5 μs and 3.0 μs, respectively, at 75 W were used. $^1$H solution-state spectra were obtained for liquid UDCH and UDSH using a BBO 800MHz S4 5mm probe. Data were obtained using a direct excitation, single pulse sequence with a 30° flip angle of 4.8us at 21W. $^1$H chemical shifts were referenced indirectly with adamantane to the $^1$H signal of tetrakis(trimethylsilyl)silane at 0.247 ppm[53]. $^{31}$P chemical shifts were referenced to the $^{31}$P signal of 85% $H_3PO_4$ at 0 ppm. $^6$Li solid-state NMR spectra were obtained using a 3.2mm HXY MAS probe with zirconia rotors sealed under Ar with Vespel caps at a spinning speed of 20 kHz. The powder samples were packed in zirconia rotors sealed with Vespel caps under argon. For the rotor synchronized spin-echo pulse sequence, 90° and 180° flip angles of 8.0 μs and 16.0 μs, respectively, at 200 W were used. $^6$Li chemical shifts were referenced to 1M LiCl at 0 ppm. Spectra destined for $^1$H quantification in the UDSH@LPSC samples were obtained using a 2.5mm HX MAS probe with zirconia rotors sealed with Vespel caps under Ar at a spinning speed of 20kHz. A larger rotor size was used in order to maximize sample content and $^1$H signal for quantification. The rotor synchronized spin-echo pulse sequence used 90° and 180° flip angles of 2.66 μs and 5.32 μs, respectively, at 100 W. $T_2$* measurements on each sample were performed to compensate for any uneven signal decay during the 50 μs echo delay. On each sample, a series of rotor synchronized spin-echos (90°-$\tau_r$-180°-$\tau_r$-acquisition) with variable echo delays was acquired and the spectra were fitted using an in-house package to obtain a $T_2$* value. To ensure accurate $^1$H quantification, a spectrum and $T_2$* measurement was also obtained on an empty rotor. Thereafter, the same measurements were repeated on a known mass of adamantane and signal contributions from the empty rotor were subtracted to calibrate $^1$H content to an integrated spectral intensity. Identical measurements were conducted on a pristine LPSC sample, and two UDSH@LPSC samples dried at 80 °C and 300 °C, respectively. For each sample, the proton content was calculated accounting for $T_2$* signal decay of each fitted component and contributions from the empty rotor. In the dried UDSH@LPSC samples, the proton content of pristine LPSC was subtracted to calculate the 1-undecanethiol content of the sample. 2D exchange spectroscopy (EXSY). LPSC and UDSH@LPSC were mixed with $Li_2ZrCl_6$ (LZC) to examine exchange of Li between the two electrolytes using 2D EXSY. The LZC electrolyte was mechanochemically synthesized using LiCl and a 10% weight excess of $ZrCl_4$, mixed with 15x 10mm $ZrO_2$ grinding media at 550 rpm for 3 hours using a Retsch PM 100 ball mill. Each of the LPSC sample was hand mixed for 15 minutes with 3 times the molar amount of LZC to match the lithium content in each environment. The resulting powder samples were packed in zirconia rotors and sealed with PTFE tape and Vespel caps under argon. 2D EXSY spectra were obtained using a 2.5mm HX MAS probe tuned to $^6$Li at an MAS speed of 30 kHz. The data were acquired using a rotor-synchronized EXSY pulse sequence with 90-degree pulses of 3.5 µs at 200W and a recycle delay of 7 seconds. Chemical shifts were referenced with respect to a 1M LiCl solution at 0 ppm. The resulting 2D spectra were analyzed using TopSpin 3.6 and fit using dmfit.

Battery Assembly: To evaluate the electrochemical performance, ASSBs Swagelok cell composed of a polyaryletheretherketone (PEEK) mold and Ti rods were assembled. In the symmetric cells, a pressure of 375 MPa was applied to compact 200 mg of solid electrolyte powder into a pellet with a diameter of 13 mm for 7 mins. Lithium metal foil, with a diameter of 1.11 cm and a thickness of 100 µm (>99.9%, MSE Supplies), was attached to both sides of the electrolyte pellet. Subsequently, the resulting Li|SSE|Li symmetric cell was sandwiched between two Ti rods. In full cells, the cathode composite was made by mixing NCM811 ($LiNi_{0.8}Co_{0.1}Mn_{0.1}O_2$, LG Energy Solution) − LPSC (>99.9%, Ampcera Inc, used as received) − vapor grown carbon fiber (>98%, Sigma Aldrich) in the weight ratio of 60:37:3 in a mortar and pestle. 100 mg of pristine LPSC and 100 mg of air exposed SSEs (specimens in Fig. 3a) were pressed at 30 MPa to form a dual-layered electrolyte pellet. Following this, a cathode composite weighing 16.2 mg was introduced onto the pristine LPSC side and pressed at 375 MPa for 7 mins. The Li metal and In metal foil, with a diameter of 1.11 cm was attached to the side of air exposed SSE and pressed at 125 MPa for 1 min (In foil, 99.99%, 0.127mm thick from Fisher Scientific). Subsequently, the resulting $Li_{0.5}In$|air exposed SSE|LPSC|cathode cell was sandwiched between two Ti rods. Symmetric and full cells were kept at 30MPa and 55MPa during testing, respectively. All cells were assembled in an argon atmosphere glovebox with oxygen and water levels below 0.5 ppm.

Electrochemical testing: Cell measurements were made on a LAND multi-channel battery testing system. The galvanostatic charge-discharge tests of symmetric cells was carried out at stepwise increasing current densities at room temperature (20±2°C) with 0.5 hour per half cycle.

The galvanostatic charge-discharge tests of full cells were conducted within the voltage range of 1.9-3.65 V at a rate of 0.15 mA cm$^{-2}$ at 60°C. The impedance measurements were conducted in the frequency range of 7 MHz to 100 mHz with an applied AC potential of 5 mV using a frequency response analyzer (Biologic workstation) at room temperature (20±2°C). The impedance measurements setup involved the use of 100 mg of electrolyte pressed and kept at 375 MPa with diameter of 13mm at room temperature (20±2°C), while the Ti rods served as blocking electrodes. All cells were tested in an argon atmosphere glovebox with oxygen and water levels below 0.5 ppm.

**Data Availability**

Source data are provided with this paper. The data that support the findings of this study are available within the article and its Supplementary Information/Source Data files.

**Acknowledgements**

This work was supported by LG Energy Solution–U.C. San Diego Frontier Research Laboratory (FRL) via the Open Innovation program. This work was partially supported by the National Science Foundation through the Materials Research Science and Engineering Center (MRSEC) at UC Santa Barbara: NSF DMR–2308708. This work was partially supported by the National Science Foundation Graduate Research Fellowship under Grant 1650114. This work was performed in part at the San Diego Nanotechnology Infrastructure (SDNI) of UCSD, a member of the National Nanotechnology Coordinated Infrastructure (NNCI), which is supported by the National Science Foundation (Grant ECCS-2025752). The authors acknowledge the use of facilities and instrumentation supported by NSF through the UC San Diego Materials Research Science and Engineering Center (UCSD MRSEC), grant # DMR-2011924. The authors acknowledge the use of facilities and instrumentation at the UC Irvine Materials Research Institute (IMRI), which is supported in part by the National Science Foundation through the UC Irvine Materials Research Science and Engineering Center (DMR-2011967). XPS work was performed using instrumentation funded in part by the National Science Foundation Major Research Instrumentation Program under grant no. CHE-1338173

**Author Contributions Statement**

P.L. conceived the idea and MC.L., J.J.H. and P.L. designed the experiments. E.S. and T.P. carried out NMR experiments with supervision from R.J.C.. MC.L., J.J.H., S.W., S.F., Z.H., E.L., Q.M. and S.Y. carried out the other experiments and measurements. K.Z. carried out the calculation. N.H., J.Z., J.O., M.S. and J.L. helped with discussion. MC.L., J.J.H., E.S., K.Z., and P.L. prepared the manuscript with contributions from all authors. P.L. supervised the research.

**Competing Interests Statement**

P.L., MC.L., J.J.H., K.Z., J.O. and M.S. have filed a US patent application based on this work. The remaining authors declare no competing interests.

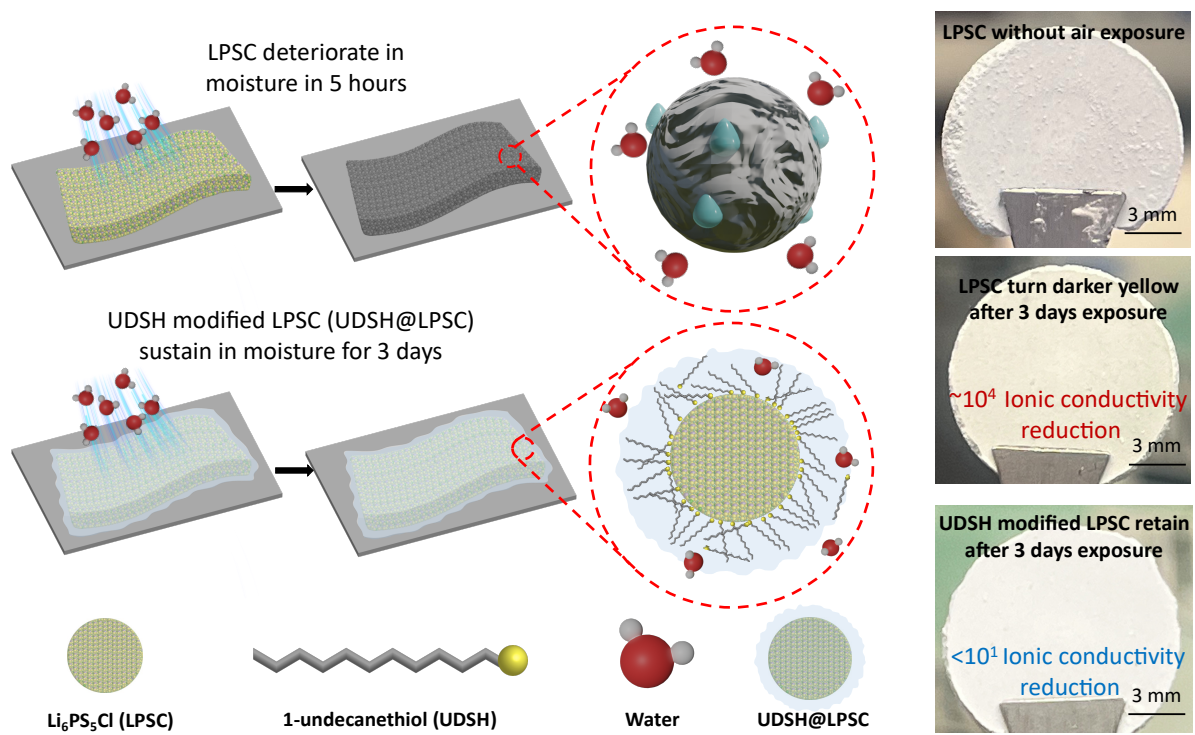

Fig. 1 Schematic illustration of the protection strategy for $Li_6PS_5Cl$ (LPSC) in humid ambient air. While LPSC undergoes hydrolysis and suffers a ~ 4 orders of magnitude reduction in ionic conductivity, a liquid coating of 1-undecanethiol (UDSH) with some of it chemically adsorbed onto the LPSC surface protects the material, which loses less than 10-fold in conductivity in 33% relative humidity (RH). UDSH forms a hydrophobic shield around the LPSC particles.

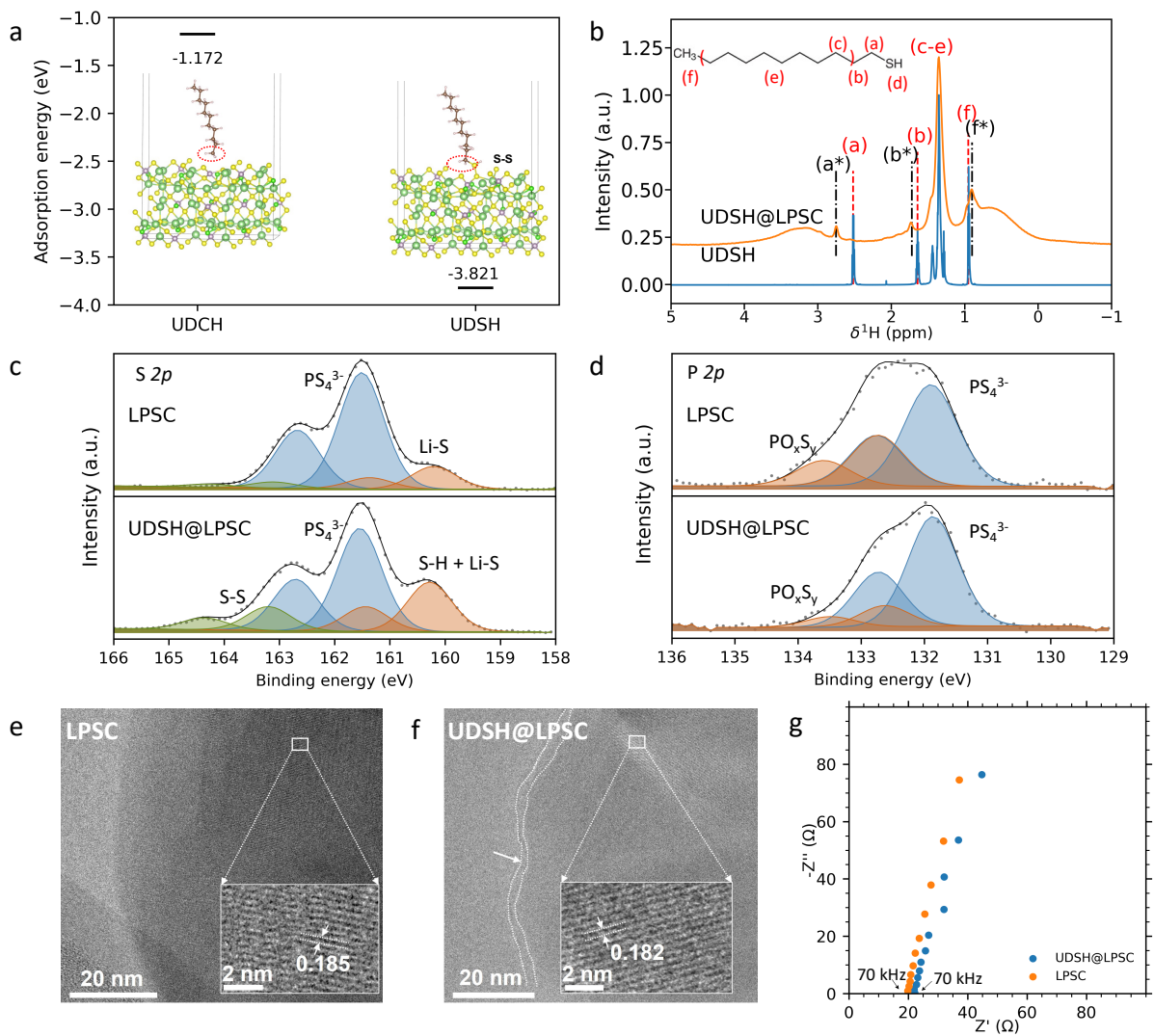

Fig. 2 Chemical compatibility between the UDSH surface modifier and LPSC. (a) Computed adsorption energies of undecane (UDCH) and UDSH onto the surface of LPSC. The six types of atoms are represented by colored spheres as follows: Li (light green), P (light purple), S (yellow), Cl (bright green), C (light Orange), and H (silver). (b) $^1$H spin echo NMR spectra collected at 18.8 T on a pure solution of UDSH and on UDSH@LPSC sample. $^1$H resonances in the UDSH@LPSC spectrum, and their corresponding positions in the molecule, are labeled (a-f) in the upper left corner. The positions of the $^1$H resonances associated with the UDSH molecule in the pure UDSH solution, and in the UDSH@LPSC sample, are indicated with a red dashed line and with a black dashed line/asterisk, respectively. S *2p* (c) and P *2p* (d) XPS spectra collected on LPSC and UDSH@LPSC. Cryo-TEM images of LPSC (e) and UDSH@LPSC (f),

visualizing the modifying layer on the LPSC surface in UDSH@LPSC. (g) Nyquist plots of LPSC and UDSH@LPSC pellets sandwiched between two titanium rods. Source data are provided as a Source Data file.

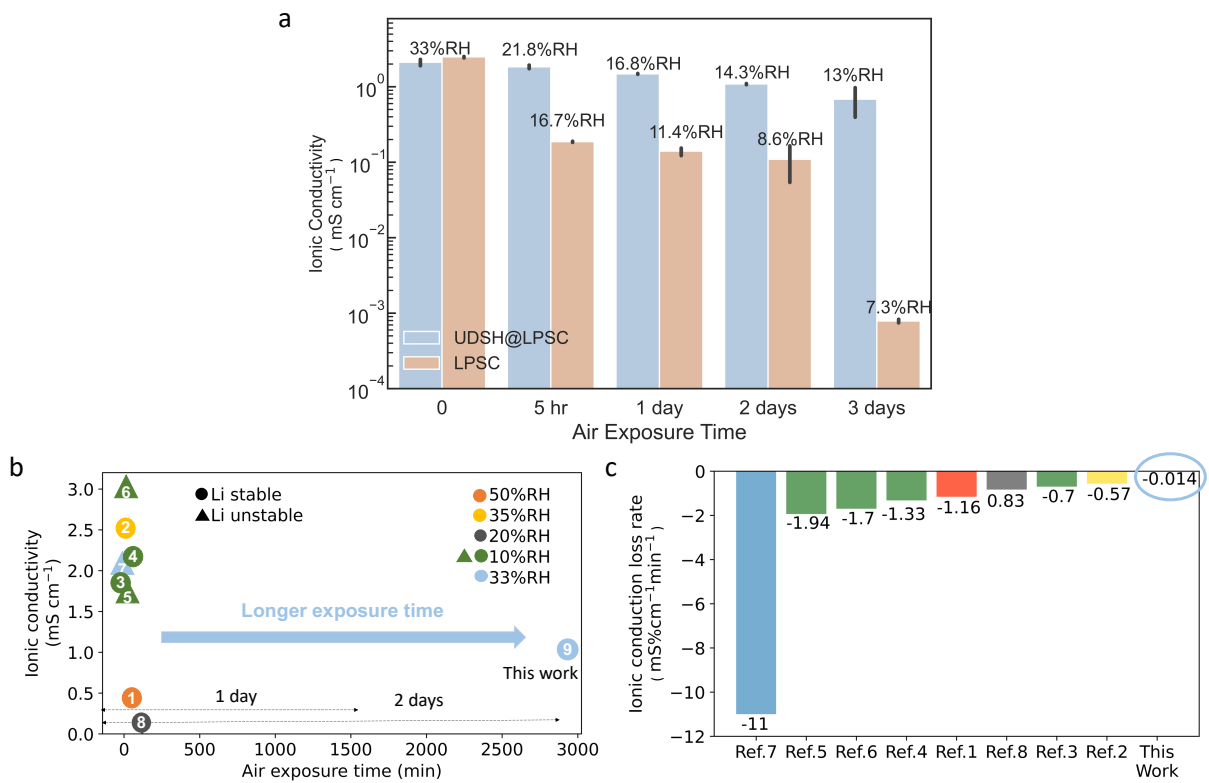

Fig. 3 Evaluation of LPSC and UDSH@LPSC upon exposure to humid ambient air. (a) Ionic conductivity of LPSC and UDSH@LPSC as a function of air exposure time. Data are presented as mean values +/- SD. (b) A comparison of ionic conductivity as a function of exposure time to humid air, highlighting the vastly superior stability of UDSH@LPSC SSE for up to two orders of magnitude longer exposure times. Samples in references 2-8 were powders tested in closed systems, similar to current work. Sample in reference 1 was a pellet tested in an open system. (c) Ionic conductivity loss rate comparison. When normalized to the exposure duration, the dramatic improvement offered by UDSH is also evident. [20,29,31,42–46]. Source data are provided as a Source Data file.

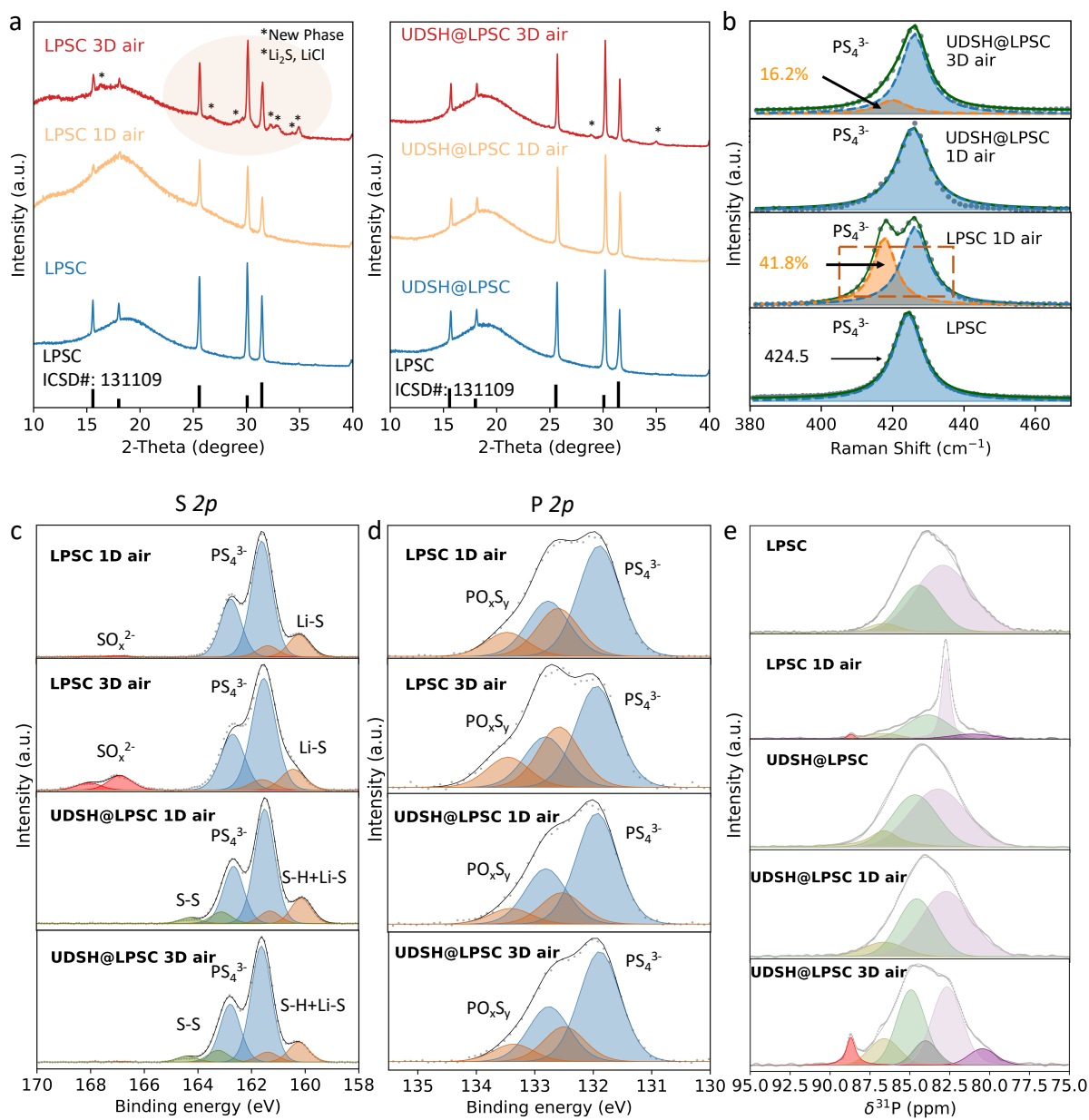

Fig. 4 Analysis of material degradation upon moisture exposure. (a) XRD patterns collected on LPSC and UDSH@LPSC exposed to humid air for up to 3 days. (b) Raman spectra of LPSC and UDSH@LPSC after 1 and 3 days of humid air exposure. (c-d) XPS and (e) $^{31}$P spin echo solid-state NMR spectra collected on LPSC and UDSH@LPSC after 1 day or 3 days of air exposure. Source data are provided as a Source Data file.

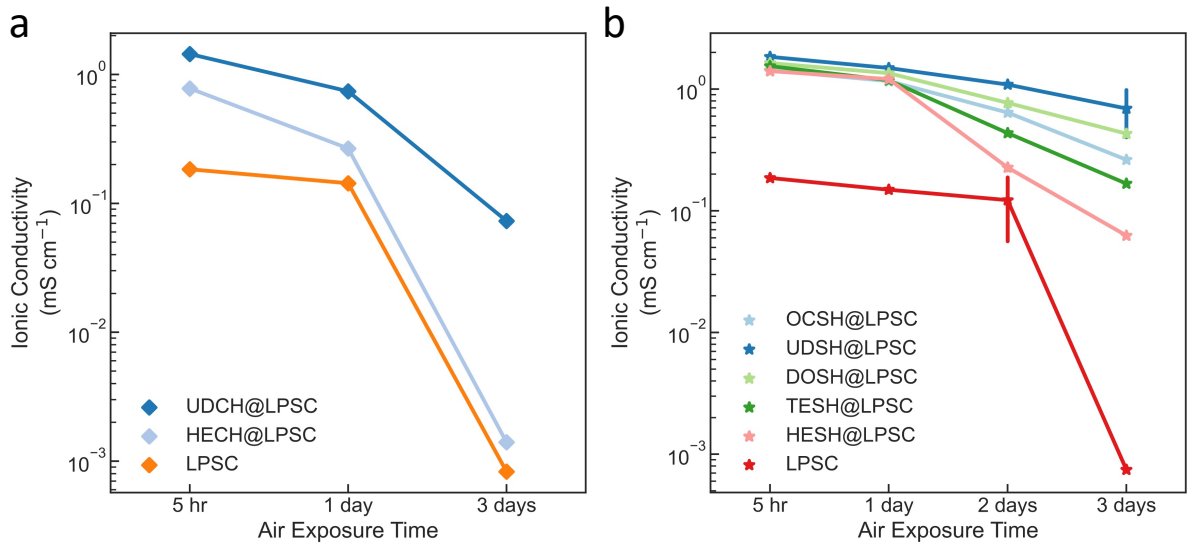

Fig. 5 Effect of functional group and chain length on LPSC protection upon moisture exposure. (a) Ionic conductivity of LPSC, UDCH@LPSC and HECH@LPSC as a function of time of exposure to humid air. (b) Ionic conductivity of LPSC, OCSH@LPSC, UDSH@LPSC, DOSH@LPSC, TESH@LPSC and HESH@LPSC as a function of time of exposure to humid air. Data are presented as mean values +/- SD. Source data are provided as a Source Data file.

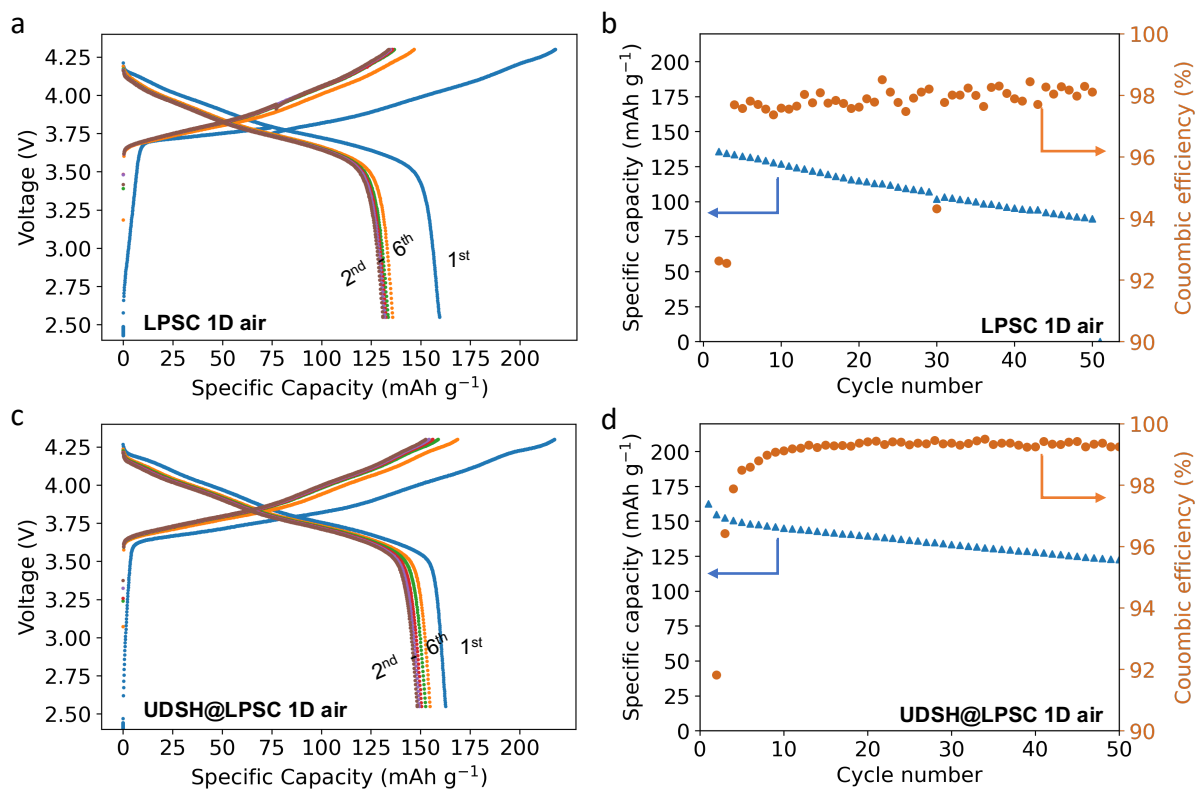

Fig. 6 Electrochemical performance of LPSC and UDSH@LPSC samples after 1 day of 33%RH air exposure. Voltage profiles of (a) an Li$_{0.5}$In| LPSC 1D air|LPSC|NCM811 composite cell, and (c) an Li$_{0.5}$In|UDSH@LPSC 1D air|LPSC|NCM811 composite cell at 60 °C and at a 0.15 mA cm$^{-2}$. (b), (d) 50 cycle capacity retention for the cells shown in (a) and (c), respectively. Source data are provided as a Source Data file.